\documentclass[letterpaper,10pt]{article} 

\usepackage{style/osameet3} 

\usepackage[colorlinks=true,bookmarks=false,citecolor=blue,urlcolor=blue]{hyperref} 
\usepackage{silence}
\usepackage{booktabs}
\usepackage{lipsum}
\usepackage[per-mode=symbol]{siunitx}
\DeclareSIUnit{\sample}{Sa}
\DeclareSIUnit{\baud}{Bd}
\DeclareSIUnit{\bit}{b}
\DeclareSIUnit{\byte}{B}

\usepackage[acronym,nomain]{glossaries}
\glsdisablehyper
\usepackage{amsmath,amssymb}
\usepackage[subrefformat=parens]{subcaption}
\usepackage[capitalize]{cleveref}
\newcommand{\SetCapsType}{normalcaps}

\makeatletter
\providecommand{\SetCapsType}{smallcaps}

\long\def\@scTrue{smallcaps}
\long\def\@scFalse{normalcaps}
\newcommand{\acroSCaps}[1]{%
 \begingroup
  \ifx\SetCapsType\@scTrue 
    \textsc{#1}%
  \else
    \MakeUppercase{#1}%
  \fi
  \endgroup
}
\makeatother

\newcommand{\nAcronym}[4][]{%
	\newacronym[#1]{#2}{#3}{#4}
}

\makeatletter
\@ifpackageloaded{babel}{%
    \newcommand{\usuk}[2]{%
        \iflanguage{USenglish}{#1}{#2}%
    }%
}{%
    \newcommand{\usuk}[2]{%
        #1%
    }%
}%
\makeatother

\usepackage[shortcuts]{extdash}
\newcommand{\qam}[1]{
    \ifglsused{QAM}%
        {#1\=/\gls{QAM}}%
        {#1\=/ary \gls{QAM}%
    }%
}%

\newcommand{\pam}[1]{
    \ifglsused{PAM}%
        {\gls{PAM}\=/#1}%
        {#1\=/ary \gls{PAM}%
    }%
}%

\nAcronym{2A8PSK}{\acroSCaps{2a8psk}}{2-ary amplitude 8-ary phaseshift keying}

\nAcronym{3CCMCF}{\acroSCaps{3cc-mcf}}{3-core coupled-core multi-core fiber}

\nAcronym{4D}{\acroSCaps{4d}}{four-dimensional}
\nAcronym{4D64PRS}{\acroSCaps{4d-64prs}}{\usuk{four-dimensional 64-ary polarization-ring-switching}{four-dimensional 64-ary polarisation-ring-switching}}

\nAcronym{5B4D2A8PSK}{\acroSCaps{5b4d-2a8psk}}{5-bit four-dimensional two-amplitude 8-ary phase-shift keying}

\nAcronym{6B4D2A8PSK}{\acroSCaps{6b4d-2a8psk}}{6-bit four-dimensional two-amplitude 8-ary phase-shift keying}

\nAcronym{8D}{\acroSCaps{8d}}{eight-dimensional}\nAcronym{8D2048PRS}{\acroSCaps{8d-2048prs}}{eight-dimensional 2048-ary polarization-ring-switching}
\nAcronym{8D2048PRST1}{\acroSCaps{8d-2048prs-t1}}{eight-dimensional 2048-ary polarization-ring-switching type 1}
\nAcronym{8D2048PRST2}{\acroSCaps{8d-2048prs-t2}}{eight-dimensional 2048-ary polarization-ring-switching type 2}

\nAcronym{ABC}{\acroSCaps{abc}}{automatic bias controller}
\nAcronym{ADC}{\acroSCaps{adc}}{analog-to-digital converter}
\nAcronym{AIR}{\acroSCaps{air}}{achievable information rate}
\nAcronym{AOM}{\acroSCaps{aom}}{acoustic optical modulator}
\nAcronym{API}{\acroSCaps{api}}{application programming interface}
\nAcronym{AR}{\acroSCaps{ar}}{achievable rate}
\nAcronym{ASE}{\acroSCaps{ase}}{amplified spontaneous emission}
\nAcronym{ASK}{\acroSCaps{ask}}{amplitude-shift keying}
\nAcronym{ASIC}{\acroSCaps{asic}}{application-specific integrated circuit}
\nAcronym{ARRWG}{\acroSCaps{a}rr\acroSCaps{wg}}{arrayed-waveguide grating}
\nAcronym{AWG}{\acroSCaps{awg}}{arbitrary-waveform generator}
\nAcronym{AWGN}{\acroSCaps{awgn}}{additive white Gaussian noise}

\nAcronym{BCH}{\acroSCaps{bch}}{Bose-Chaudhuri-Hocquenghem}
\nAcronym{BER}{\acroSCaps{ber}}{bit error rate}
\nAcronym{BICM}{\acroSCaps{bicm}}{bit-interleaved coded modulation}
\nAcronym{BMD}{\acroSCaps{bmd}}{bit-metric decoding}
\nAcronym{BPD}{\acroSCaps{bpd}}{balanced photo-diode}
\nAcronym{BPF}{\acroSCaps{bpf}}{bandpass filter}
\nAcronym{BPS}{\acroSCaps{bps}}{blind phase search}
\nAcronym{BPSK}{\acroSCaps{bpsk}}{binary phase-shift keying}
\nAcronym{BRGC}{\acroSCaps{brgc}}{binary reflected Gray code}
\nAcronym{BTB}{\acroSCaps{btb}}{back-to-back}

\nAcronym{CAGR}{\acroSCaps{cagr}}{compound annual growth rate}
\nAcronym{CCDM}{\acroSCaps{ccdm}}{constant composition distribution matching}
\nAcronym{CCF}{\acroSCaps{ccf}}{\usuk{coupled-core fiber}{coupled-core fibre}}
\nAcronym{CD}{\acroSCaps{cd}}{chromatic dispersion}
\nAcronym{CIR}{\acroSCaps{cir}}{channel impulse response}
\nAcronym{CMA}{\acroSCaps{cma}}{constant modulus algorithm}
\nAcronym{CMUX}{\acroSCaps{cmux}}{core multiplexer}
\nAcronym{ChUT}{\acroSCaps{chut}}{channel under test}
\nAcronym{CUT}{\acroSCaps{cut}}{channel under test}
\nAcronym{CPE}{\acroSCaps{cpe}}{carrier phase estimation}
\nAcronym{CPU}{\acroSCaps{cpu}}{central processing unit}
\nAcronym{CSPR}{\acroSCaps{cspr}}{carrier-to-signal power ratio}
\nAcronym{CW}{\acroSCaps{cw}}{continuous wave}

\nAcronym{DA}{\acroSCaps{da}}{driver amplifier}
\nAcronym{DAC}{\acroSCaps{dac}}{digital-to-analog converter}
\nAcronym{DCF}{\acroSCaps{dcf}}{\usuk{dispersion compensated fiber}{dispersion compensated fibre}}
\nAcronym{DCI}{\acroSCaps{dci}}{data center interconnect}
\nAcronym{DDLMS}{\acroSCaps{dd-lms}}{decision-directed least mean square}
\nAcronym{DFB}{\acroSCaps{dfb}}{distributed feedback}
\nAcronym{DGD}{\acroSCaps{dgd}}{differential group delay}
\nAcronym{DH}{\acroSCaps{dh}}{digital holography}
\nAcronym{DM}{\acroSCaps{dm}}{distribution matcher}
\nAcronym{DMA}{\acroSCaps{dma}}{direct memory access}
\nAcronym{DMD}{\acroSCaps{dmd}}{differential mode delay}
\nAcronym{DMG}{\acroSCaps{dmg}}{differential modal gain}
\nAcronym{DMGD}{\acroSCaps{dmgd}}{differential mode group delay}
\nAcronym{DML}{\acroSCaps{dml}}{directly-modulated laser}
\nAcronym{DP}{\acroSCaps{dp}}{\usuk{dual-polarization}{dual-polarisation}}
\nAcronym{DPC}{\acroSCaps{dpc}}{digital pre-compensation}
\nAcronym{DPE}{\acroSCaps{dpe}}{digital pre-emphasis}
\nAcronym{DPIQ}{\acroSCaps{dp-iqm}}{\usuk{dual-polarization IQ-modulator}{dual-polarisation IQ-modulator}}
\nAcronym{DPLL}{\acroSCaps{dpll}}{digital phase-locked loop}
\nAcronym{DQPSK}{\acroSCaps{dqpsk}}{differential quaternary phase-shift-keying}
\nAcronym{DRA}{\acroSCaps{dra}}{distributed Raman amplifier}
\nAcronym{DRE}{\acroSCaps{dre}}{digital resolution enhancer}
\nAcronym{DSF}{\acroSCaps{dsf}}{\usuk{dispersion-shifted fiber}{dispersion-shifted fibre}}
\nAcronym{DSO}{\acroSCaps{dso}}{digital sampling oscilloscope}
\nAcronym{DSP}{\acroSCaps{dsp}}{digital signal processing}
\nAcronym{DUT}{\acroSCaps{dut}}{device-under-test}
\nAcronym{DWDM}{\acroSCaps{dwdm}}{dense wavelength-division multiplexing}

\nAcronym{EAM}{\acroSCaps{eam}}{electro-absorption modulator}
\nAcronym{ECL}{\acroSCaps{ecl}}{external cavity laser}
\nAcronym{ED}{\acroSCaps{ed}}{Eucledian distance}
\nAcronym{EDFA}{\acroSCaps{edfa}}{\usuk{erbium-doped fiber amplifier}{erbium-doped fibre amplifier}}
\nAcronym{ENOB}{\acroSCaps{enob}}{effective number of bits}
\nAcronym{ESS}{\acroSCaps{ess}}{enumerative sphere shaping}

\nAcronym{FD}{\acroSCaps{fd}}{frequency domain}
\nAcronym{FDE}{\acroSCaps{fde}}{\usuk{frequency domain equalizer}{frequency domain equaliser}}
\nAcronym{FEC}{\acroSCaps{fec}}{forward error correction}
\nAcronym{FFT}{\acroSCaps{fft}}{fast Fourier transform}
\nAcronym{FIR}{\acroSCaps{fir}}{finite impulse response}
\nAcronym{FMF}{\acroSCaps{fmf}}{\usuk{few-mode fiber}{few-mode fibre}}
\nAcronym[plural=FM-MCF, firstplural=\usuk{few-mode multi-core fibers}{few-mode multi-core fibres}]{FM-MCF}{\acroSCaps{fm-mcf}}{\usuk{few-mode multi-core fiber}{few-mode multi-core fibre}}
\nAcronym{FPGA}{\acroSCaps{fpga}}{field-programmable gate array}
\nAcronym{FWM}{\acroSCaps{fwm}}{four-wave mixing}

\nAcronym{GD}{\acroSCaps{gd}}{group delay}
\nAcronym{GI}{\acroSCaps{gi}}{graded-index}
\nAcronym{GFF}{\acroSCaps{gff}}{gain flattening filter}
\nAcronym{GIMMF}{\acroSCaps{gi-mmf}}{\usuk{graded-index multi-mode fiber}{graded-index multi-mode fibre}}
\nAcronym{GMI}{\acroSCaps{gmi}}{\usuk{generalized mutual information}{generalised mutual information}}
\nAcronym{GNSE}{\acroSCaps{gnse}}{generalized nonlinear Schr\"{o}dinger equation}
\nAcronym{GPU}{\acroSCaps{gpu}}{graphics processing unit}
\nAcronym{GS}{\acroSCaps{gs}}{geometric shaping}
\nAcronym{GV}{\acroSCaps{gv}}{group-velocity}
\nAcronym{GVD}{\acroSCaps{gvd}}{group-velocity dispersion}
\nAcronym{GPIO}{\acroSCaps{gpio}}{general purpose input output}

\nAcronym{HDFEC}{\acroSCaps{hd-fec}}{hard decision forward error correction}

\nAcronym[plural=IL, firstplural=insertion losses (\textsc{il})]{IL}{\acroSCaps{il}}{insertion loss}
\nAcronym{IFFT}{\acroSCaps{ifft}}{inverse fast Fourier transform}
\nAcronym{IMDD}{\acroSCaps{im/dd}}{intensity-modulation direct-detection}
\nAcronym{IQM}{\acroSCaps{iqm}}{in-phase and quadrature modulator}
\nAcronym{ISI}{\acroSCaps{isi}}{intersymbol interference}

\nAcronym{KK}{\acroSCaps{kk}}{Kramers-Kronig}

\nAcronym{LCOS}{\acroSCaps{LCoS}}{liquid crystal on silicon}
\nAcronym{LDPC}{\acroSCaps{ldpc}}{low-density parity-check}
\nAcronym{LEAF}{\acroSCaps{leaf}}{\usuk{large effective area fiber}{large effective area fibre}}
\nAcronym{LFSR}{\acroSCaps{lfsr}}{linear-feedback shift register}
\nAcronym{LMS}{\acroSCaps{lms}}{least means square}
\nAcronym{LLR}{\acroSCaps{llr}}{log-likelihood ratio}
\nAcronym{LO}{\acroSCaps{lo}}{local oscillator}
\nAcronym{LP}{\acroSCaps{lp}}{\usuk{linearly polarized}{linearly polarised}}
\nAcronym{LSPS}{\acroSCaps{lsps}}{\usuk{loop-synchronized polarization scrambler}{loop-synchronised polarisation scrambler}}
\nAcronym{LUT}{\acroSCaps{lut}}{lookup table}

\nAcronym{MB}{\acroSCaps{mb}}{Maxwell-Bolzmann}
\nAcronym{MCF}{\acroSCaps{mcf}}{\usuk{multi-core fiber}{multi-core fibre}}
\nAcronym{MDG}{\acroSCaps{mdg}}{mode dependent gain}
\nAcronym[plural=MDL, firstplural=mode-dependent losses (\textsc{mdl})]{MDL}{\acroSCaps{mdl}}{mode-dependent loss}
\nAcronym{MDM}{\acroSCaps{mdm}}{mode division multiplexing}
\nAcronym{MEMS}{\acroSCaps{mems}}{micro-electro-mechanical systems}
\nAcronym{MF}{\acroSCaps{mf}}{matched filter}
\nAcronym{MI}{\acroSCaps{mi}}{mutual information}
\nAcronym{MIMO}{\acroSCaps{mimo}}{multiple-input multiple-output}
\nAcronym{ML}{\acroSCaps{ml}}{machine learning}
\nAcronym{MMA}{\acroSCaps{mma}}{multi-modulus algorithm}
\nAcronym{MMF}{\acroSCaps{mmf}}{\usuk{multi-mode fiber}{multi-mode fibre}}
\nAcronym{MMSE}{\acroSCaps{mmse}}{minimum mean squared error}
\nAcronym{MP}{\acroSCaps{mp}}{minimum phase}
\nAcronym{MPLC}{\acroSCaps{mplc}}{multi-plane light converter}
\nAcronym{MSE}{\acroSCaps{mse}}{mean squared error}
\nAcronym{MUX}{\acroSCaps{mux}}{multiplexer}
\nAcronym{MZM}{\acroSCaps{mzm}}{Mach-Zehnder modulator}
\nAcronym{MZI}{\acroSCaps{mzi}}{Mach-Zehnder interferometer}

\nAcronym{NA}{\acroSCaps{na}}{numerical aperture}
\nAcronym{NF}{\acroSCaps{nf}}{noise figure}
\nAcronym{NGMI}{\acroSCaps{ngmi}}{\usuk{normalized generalized mutual information}{normalised generalised mutual information}}
\nAcronym{NLSE}{\acroSCaps{nlse}}{nonlinear Schr\"{o}ding equation}
\nAcronym{NIR}{\acroSCaps{nir}}{near-infrared}

\nAcronym{OBTB}{\acroSCaps{obtb}}{optical back-to-back}
\nAcronym{ODE}{\acroSCaps{ode}}{ordinary differential equation}
\nAcronym{ODL}{\acroSCaps{odl}}{optical delay line}
\nAcronym{OFC}{\acroSCaps{ofc}}{Optical Fiber Communications Conference}
\nAcronym{OFDR}{\acroSCaps{ofdr}}{optical frequency-domain reflectometer}
\nAcronym{OFDM}{\acroSCaps{ofdm}}{orthogonal frequency division multiplexing}
\nAcronym{OMFT}{\acroSCaps{omft}}{optical multi-format transmitter}
\nAcronym{OOK}{\acroSCaps{ook}}{on-off keying}
\nAcronym{OPLL}{\acroSCaps{opll}}{optical phase-locked loop}
\nAcronym{OSA}{\acroSCaps{osa}}{\usuk{optical spectrum analyzer}{optical spectrum analyser}}
\nAcronym{OSNR}{\acroSCaps{osnr}}{optical signal-to-noise ratio}
\nAcronym{OTDR}{\acroSCaps{otdr}}{optical time-domain reflectometer}
\nAcronym{OTF}{\acroSCaps{otf}}{optical tunable filter}
\nAcronym{OVNA}{\acroSCaps{ovna}}{\usuk{optical vector network analyzer}{optical vector network analyser}}

\nAcronym{PAM}{\acroSCaps{pam}}{pulse-amplitude modulation}
\nAcronym{PAS}{\acroSCaps{pas}}{probabilistic amplitude shaping}
\nAcronym{PAPR}{\acroSCaps{papr}}{peak-to-avarage power ratio}
\nAcronym{PBC}{\acroSCaps{pbc}}{\usuk{polarization beam combiner}{polarisation beam combiner}}
\nAcronym{PBS}{\acroSCaps{pbs}}{polarization beam splitter}
\nAcronym{PCVD}{\acroSCaps{pcvd}}{plasma chemical vapor depostion}
\nAcronym{PD}{\acroSCaps{pd}}{photodiode}
\nAcronym{PDL}{\acroSCaps{pdl}}{polarization-dependent loss}
\nAcronym{PDM}{\acroSCaps{pdm}}{\usuk{polarization-division multiplexing}{polarisation-division multiplexing}}
\nAcronym{PL}{\acroSCaps{pl}}{photonic lantern}
\nAcronym{PMBPSK}{\acroSCaps{pm-bpsk}}{polarization-multiplexed binary phase-shift-keying}
\nAcronym{PMQPSK}{\acroSCaps{pm-qpsk}}{polarization-multiplexed quaternary phase-shift-keying}
\nAcronym{PM8QAM}{\acroSCaps{pm-8qam}}{polarization-multiplexed 8-ary quadrature amplitude modulation}
\nAcronym{PMD}{\acroSCaps{pmd}}{polarization mode dispersion}
\nAcronym{PNOB}{\acroSCaps{pnob}}{physical number of bits}
\nAcronym{PON}{\acroSCaps{pon}}{passive-optical network}
\nAcronym{PRBS}{\acroSCaps{prbs}}{pseudorandom bit sequence}
\nAcronym{PS}{\acroSCaps{ps}}{probabilistic shaping}
\nAcronym{PSK}{\acroSCaps{psk}}{phase-shift keying}
\nAcronym{PSP}{\acroSCaps{psp}}{\usuk{principal states of polarization}{principal states of polarisation}}

\nAcronym{QAM}{\acroSCaps{qam}}{quadrature amplitude modulation}
\nAcronym{QPSK}{\acroSCaps{qpsk}}{quaternary phase-shift-keying}

\nAcronym{RAM}{\acroSCaps{ram}}{random-access memory}
\nAcronym{RF}{\acroSCaps{rf}}{radio frequency}
\nAcronym{RRC}{\acroSCaps{rrc}}{root-raised-cosine}
\nAcronym{ROADM}{\acroSCaps{roadm}}{reconfigurable optical add-drop multiplexer}
\nAcronym{ROI}{\acroSCaps{roi}}{region of interest}

\nAcronym{SA}{\acroSCaps{sa}}{simulated annealing}
\nAcronym{SamPerSym}{\acroSCaps{sps}}{samples per symbol}
\nAcronym{SBS}{\acroSCaps{sbs}}{stimulated Brillouin scattering}
\nAcronym{SDFEC}{\acroSCaps{sd-fec}}{soft decision forward error correction}
\nAcronym{SDM}{\acroSCaps{sdm}}{space-division multiplexing}
\nAcronym{SE}{\acroSCaps{se}}{spectral efficiency}
\nAcronym{SER}{\acroSCaps{ser}}{symbol error rate}
\nAcronym{SI}{\acroSCaps{si}}{step index}
\nAcronym{SLM}{\acroSCaps{slm}}{spatial light modulator}
\nAcronym{SMF}{\acroSCaps{smf}}{\usuk{single-mode fiber}{single-mode fibre}}
\nAcronym{SNR}{\acroSCaps{snr}}{signal-to-noise ratio}
\nAcronym{SOA}{\acroSCaps{soa}}{semiconductor optical amplifier}
\nAcronym[firstplural=\usuk{states of polarization (\acroSCaps{sop})}{states of polarisation (\acroSCaps{sop})}]{SOP}{\acroSCaps{sop}}{\usuk{state of polarization}{state of polarisation}}
\nAcronym{SPM}{\acroSCaps{spm}}{self-phase modulation}
\nAcronym{SPS}{\acroSCaps{sps}}{\usuk{synchronized polarization scrambler}{synchronised polarisation scrambler}}
\nAcronym{SRS}{\acroSCaps{srs}}{stimulated Raman scattering}
\nAcronym{SSBI}{\acroSCaps{ssbi}}{signal-signal beat interference}
\nAcronym{SSFM}{\acroSCaps{ssfm}}{split-step Fourier method}
\nAcronym{SSMF}{\acroSCaps{ssmf}}{\usuk{standard single-mode fiber}{standard single-mode fibre}}
\nAcronym{STL}{\acroSCaps{stl}}{swept tunable laser}
\nAcronym{SVD}{\acroSCaps{svd}}{singular value decomposition}
\nAcronym{SWI}{\acroSCaps{swi}}{swept wavelength interferometry}

\nAcronym{TD}{\acroSCaps{td}}{time domain}
\nAcronym{TDE}{\acroSCaps{tde}}{\usuk{time domain equalizer}{time domain equaliser}}
\nAcronym{TDMSDM}{\acroSCaps{tdm-sdm}}{time-domain multiplexed space-division multiplexing}
\nAcronym{TE}{\acroSCaps{te}}{transverse electric}
\nAcronym{TIA}{\acroSCaps{tia}}{trans-impedance amplifier}
\nAcronym{TM}{\acroSCaps{TM}}{transverse magnetic}
\nAcronym{TH4D}{\acroSCaps{th-4d}}{time domain hybrid four-dimensional}
\nAcronym{TH4D2A8PSK}{\acroSCaps{th-4d-2a8psk}}{time domain hybrid four-dimensional two-amplitude eight-phase shift keying}

\nAcronym{VCSEL}{\acroSCaps{vcsel}}{vertical-cavity surface emitting laser}
\nAcronym{VOA}{\acroSCaps{voa}}{variable optical attenuator}

\nAcronym{WDM}{\acroSCaps{wdm}}{wavelength-division multiplexing}
\nAcronym{WGA}{\acroSCaps{wga}}{weakly guiding approximation}
\nAcronym{WGN}{\acroSCaps{wgn}}{white Gaussian noise}
\nAcronym{WSS}{\acroSCaps{wss}}{wavelength selective switch}

\nAcronym{XPM}{\acroSCaps{xpm}}{cross-phase modulation}
\nAcronym{XT}{\acroSCaps{xt}}{cross-talk}

\nAcronym{3DWG}{\acroSCaps{3dwg}}{3D-waveguide}

\usepackage{todonotes}

\begin{document}

\title{\acrlong{KK} Receiver Combined With Digital Resolution Enhancer}

\author{Menno van den Hout, Sjoerd van der Heide and Chigo Okonkwo}
\address{%
High-Capacity Optical Transmission Laboratory, Eindhoven University of Technology,\\PO Box 513, 5600 MB, Eindhoven, The Netherlands.}
\email{m.v.d.hout@tue.nl\vspace{-5mm}}

\copyrightyear{2021}

\begin{abstract} 
A Kramers-Kronig receiver with a continuous wave tone added digitally at the transmitter is combined with a digital resolution enhancer to limit the increase in transmitter quantization noise. Performance increase is demonstrated, as well as the ability to reduce the number of bits in the digital-to-analog converter. \vspace{-5mm}
\end{abstract}

%


\section{Introduction}
Recently, the \gls{KK} receiver \cite{Mecozzi:16} structure has been investigated as an alternative to a standard coherent receiver comprising a 90 degree hybrid. The \gls{KK} receiver is able to reconstruct the amplitude and phase of an optical signal by combining the signal with a \gls{CW} tone at the edge of the signal spectrum and detecting the amplitude of the beating between signal and \gls{CW} using a single \gls{PD}. Without the need for an optical hybrid and using only a single photodiode, the \gls{KK} receiver allows the reduction in hardware complexity and costs, which makes it an attractive receiver structure for cost sensitive short reach links. The \gls{KK} receiver has been proven to be compatible with polarization-~\cite{AntonelliPolmuxKK,Chen2019} and space-division multiplexing~\cite{VanderHeide2018,Luis2018KK}.

The \gls{CW} tone needed for signal reconstruction can be added optically at the transmitter or receiver, or can be added to the transmitted signal by modulating the same \gls{CW} which is used to modulate the signal. The \gls{CW} tone is added either electrically, by combining the \gls{RF} signals from the \glspl{DAC} with an electrical \gls{CW} tone, or digitally by adding the \gls{CW} tone before digital-to-analog conversion. In terms of hardware complexity, adding the \gls{CW} tone digitally is the most favorable, as it does not need an extra laser or \gls{RF} power combiners. However, adding the tone digitally increases quantization noise, limiting the performance.

\Gls{DRE} has been proposed in \cite{Yoffe2019_2} to increase the effective resolution of \glspl{DAC} by shaping the quantization noise added by the \glspl{DAC}, 
and has been previously demonstrated for transmitting high-cardinality signal constellations with low resolution \glspl{DAC} with a limited penalty \cite{Yoffe2019_2,Hout2010JLT}.

In this work, we investigate the ability of \gls{DRE} to mitigate the increase of quantization loss due to the digitally added \gls{CW} tone needed for the \gls{KK} receiver. For transmission over \SI{50}{\km} of \gls{SSMF} using a 6 bit \gls{DAC}, \Gls{NGMI} gains up to {0.013} are achieved by using \gls{DRE}, corresponding to a {\SI{2}{\dB}} reduction in required \gls{OSNR}. Furthermore, the number of bits of the \gls{DAC} is reduced to 4 and 5 bits with limited performance penalty.

\section{Adding \gls{CW} tone digitally at transmitter}
\begin{figure*}
    \centering
    \begin{subfigure}[t]{0.33\textwidth}
        \centering
        \includegraphics[width=\textwidth]{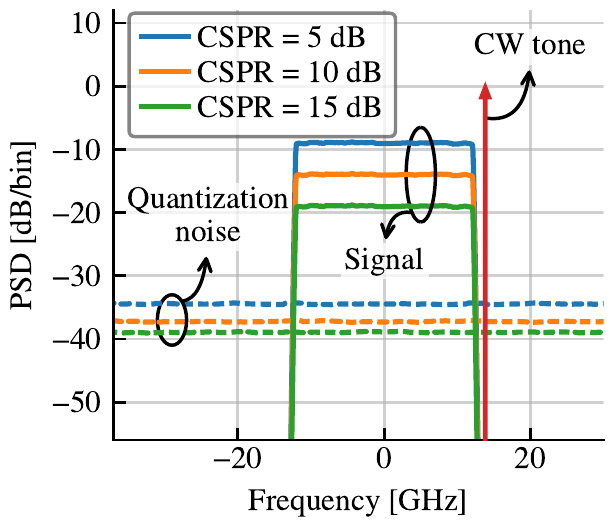}
        \vspace*{-5mm}\caption{}\label{fig:qnoise_wo_dre}
    \end{subfigure}%
    \begin{subfigure}[t]{.33\textwidth}
        \centering
        \includegraphics[width=\textwidth]{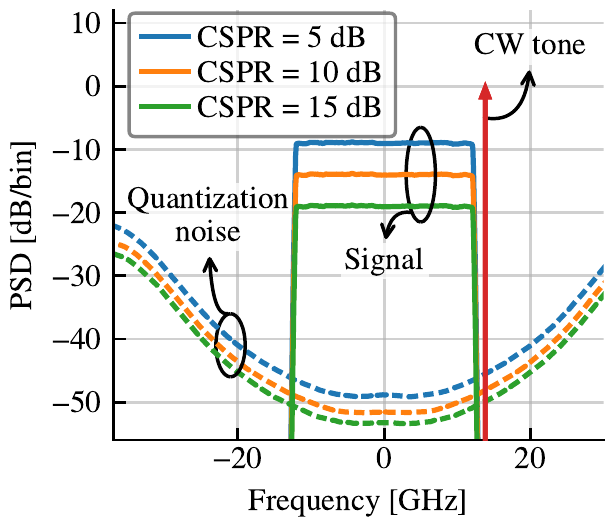}
        \vspace*{-5mm}\caption{}\label{fig:qnoise_w_dre}
    \end{subfigure}%
    \begin{subfigure}[t]{.33\textwidth}
        \centering
        \includegraphics[width=\textwidth]{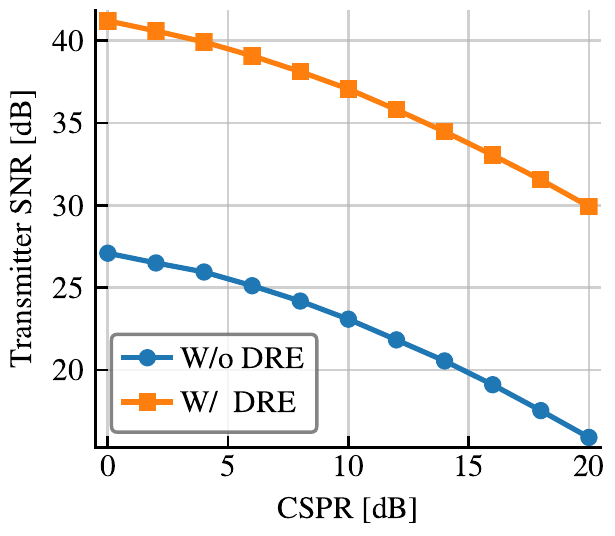}
        \vspace*{-5mm}\caption{}\label{fig:cspr_vs_tx_snr}
    \end{subfigure}%
    \vspace{-5mm}
    \caption{Spectrum of quantization noise without \textbf{\subref{fig:qnoise_wo_dre}} and with \textbf{\subref{fig:qnoise_w_dre}} DRE, and CSPR vs Transmitter SNR in \textbf{\subref{fig:cspr_vs_tx_snr}}}
    \label{fig:qnoise}
\end{figure*}

In order to perform the \gls{KK} algorithm at the receiver, the \gls{CW} tone needs to be of sufficient high power to fulfill the minimum phase condition \cite{Mecozzi:16}. Hence, the \gls{CSPR}, given by $CSPR = P_{\text{CW}}/P_{\text{sig}}$, needs to be sufficiently high as well. In the case of digitally adding the \gls{CW} tone, optimizing the \gls{CSPR} is important as there is a trade-off between the quality of the transmitted and received signal. High \glspl{CSPR} will result in better signal reconstruction at the receiver, but also result in increased quantization noise at the transmitter. The increased quantization noise is simulated and shown in \cref{fig:qnoise_wo_dre}, where the spectrum of the quantization noise resulting from a 5 bit \gls{DAC} is shown for different \glspl{CSPR}, as well as the corresponding transmitted signal and the \gls{CW}. In \cref{fig:cspr_vs_tx_snr} the \gls{CSPR} vs transmitter \gls{SNR} inside the signal band is plotted, where only quantization noise is assumed as noise source. As can be seen, increasing the \gls{CSPR} reduces the \gls{SNR}.

Using the \gls{DRE}, the quantization noise from the \gls{DAC} in the frequency region of the signal and \gls{CW} tone is reduced. This is shown in \cref{fig:qnoise_w_dre,fig:cspr_vs_tx_snr}, where 5 taps were used to model the channel and the number of soft quantization levels is 3 (see \cite{Yoffe2019_2}). Next to reducing the overall quantization noise, it is also expected that using \gls{DRE} will increase the optimum \gls{CSPR}, as the quantization noise is lower. Higher \gls{CSPR} results in better signal reconstruction, as explained above.

\section{Experimental setup}
\begin{figure}
    \centering
    \vspace{-4mm}
    \includegraphics[width=1\textwidth]{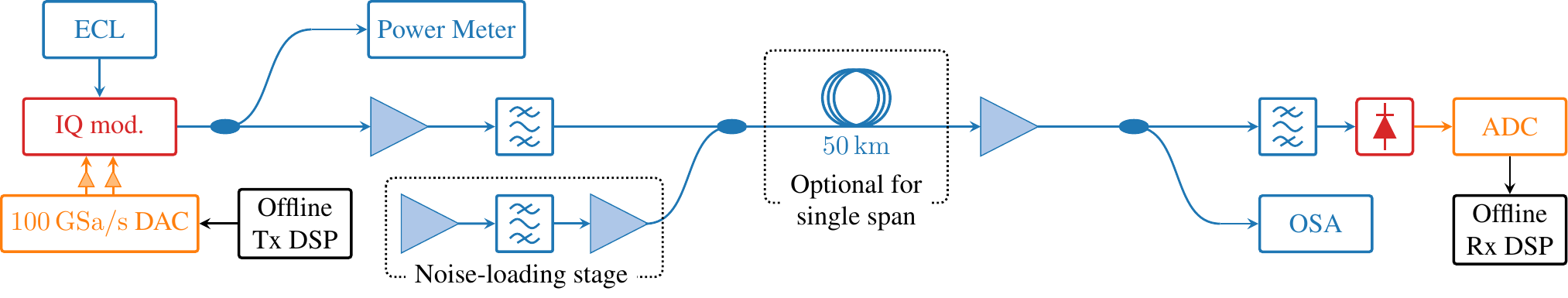}
    \vspace{-7mm}
    \caption{Used experimental setup. A tone from an \acrfull{ECL} is modulated by a modulator which modulates both the signal and the \gls{CW} tone needed for \gls{KK} detection. A single photodiode is used at the receiver.}
    \label{fig:setup}\vspace{-7mm}
\end{figure}

\begin{figure}[b]
    \centering
    \begin{subfigure}[t]{0.4\textwidth}
        \centering
        \includegraphics[width=\textwidth]{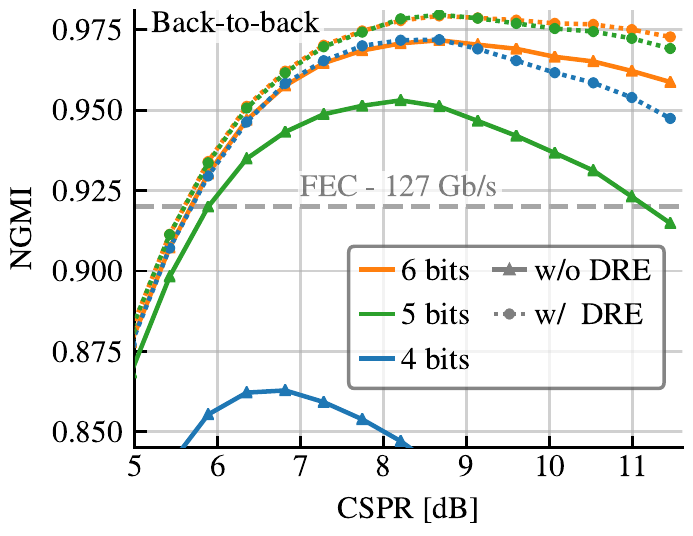}\vspace*{-2mm}
        \caption{}\label{fig:b2b_cspr}
    \end{subfigure}%
    \begin{subfigure}[t]{.4\textwidth}
        \centering
        \includegraphics[width=\textwidth]{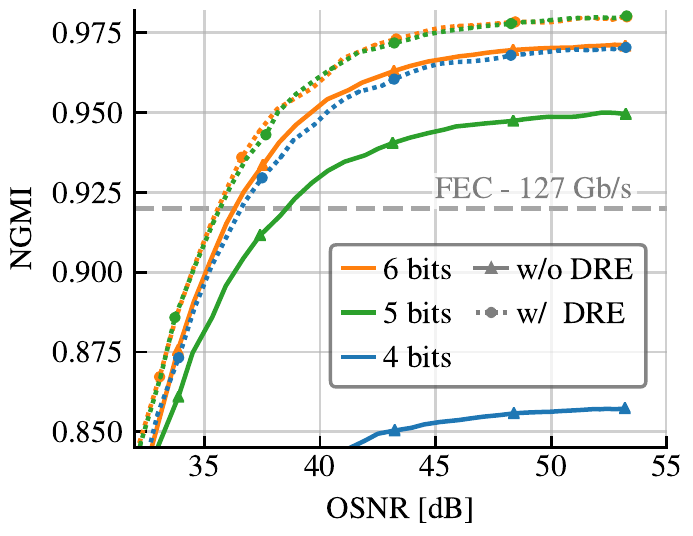}\vspace*{-2mm}
        \caption{}\label{fig:b2b_osnr}
    \end{subfigure}%
  \begin{subtable}[b]{0.2\textwidth}
      \centering
      \footnotesize
            \setlength\tabcolsep{1.5pt}
            \begin{tabular}{p{1.0cm}cc}\toprule
                 & w/o DRE & w/ DRE \\\midrule
                \multicolumn{1}{l}{\emph{6 bits}}& & \\
                \multicolumn{1}{r}{NGMI} & 0.972 & 0.980 \\
                \multicolumn{1}{r}{CSPR}&  8.7 & 8.7 \\
                \multicolumn{1}{l}{\emph{5 bits}}& & \\
                \multicolumn{1}{r}{NGMI} & 0.953 & 0.980 \\
                \multicolumn{1}{r}{CSPR}&  8.2 & 8.7 \\
                \multicolumn{1}{l}{\emph{4 bits}}& & \\
                \multicolumn{1}{r}{NGMI} & 0.862 & 0.972 \\
                \multicolumn{1}{r}{CSPR}&  6.8 & 8.4
                \\\bottomrule
            \end{tabular}
      \caption{}
      \label{table:bb}
  \end{subtable}
      \vspace{-5mm}
    \caption{Experimental results for \glsfirst{OBTB} transmission, showing in \textbf{\subref{fig:b2b_cspr}} the \gls{NGMI} vs \gls{CSPR} and in \textbf{\subref{fig:b2b_osnr}} \gls{NGMI} vs \gls{OSNR} at the optimal \gls{CSPR}. \textbf{\subref{table:bb}} summarizes the maximum \gls{NGMI} and optimum \gls{CSPR}.}
\label{fig:bb}
\end{figure}

\label{sec:setup}
In \cref{fig:setup}, the experimental setup used to validate the \gls{KK} receiver combined with \gls{DRE} is depicted. 64-QAM symbols are generated at \SI{25}{\giga\baud}, \acrlong{RRC} shaped with $\beta=0.01$ and a \gls{CW} tone is added with a \SI{13.9}{\giga\hertz} offset, followed by the \gls{DRE}. The samples are uploaded to a 6 bit \gls{DAC} operating at \SI{100}{\giga\sample\per\second}, which is connected via \glspl{DA} to a single-polarization \gls{IQM}. The \SI{193.4}{\tera\hertz} tone produced by an \gls{ECL} is modulated by the \gls{IQM} producing a \gls{MP} signal, after which it filtered by a \SI{40}{\giga\hertz} optical band-pass filter. The resulting signal is subsequently amplified and transmitted over \SI{50}{\kilo\meter} of \gls{SSMF}. At the receiver, the signal is amplified, filtered by a \gls{WSS} and detected using a single AC coupled photodiode. The resulting electrical signal is digitized by an \SI{80}{\giga\sample\per\second} \gls{ADC}. The DC bias needed for AC-coupled \gls{KK} receivers is optimized\cite{Luis:20}, after which the \gls{KK} algorithm is applied, followed by \gls{MIMO} equalization updated by a fully supervised \gls{LMS} algorithm, Finally, \gls{NGMI} evaluation is performed.

For \gls{OBTB} measurements, the \SI{50}{\km} fiber is removed. At the transmitter, noise-loading is achieved by filtering the \acrlong{ASE} of an \gls{EDFA} with a band-pass filter of \SI{250}{\giga\hertz} centered around \SI{193.4}{\tera\hertz} using a \gls{WSS}. The resulting noise band is amplified depending on the desired \gls{OSNR}, and coupled with the signal. An \gls{OSA} is used for \gls{OSNR} evaluation.

In order to measure the \gls{CSPR} of the transmitted signal, an optical power meter is placed after the \gls{IQM}. Both the signal and the \gls{CW} tone are uploaded to the \gls{DAC} and their corresponding optical powers are measured, with the \gls{CSPR} resulting from the ratio between the two powers.

\section{Experimental results}

\begin{figure}
    \centering
    \begin{subfigure}[t]{0.4\textwidth}
        \centering
        \includegraphics[width=\textwidth]{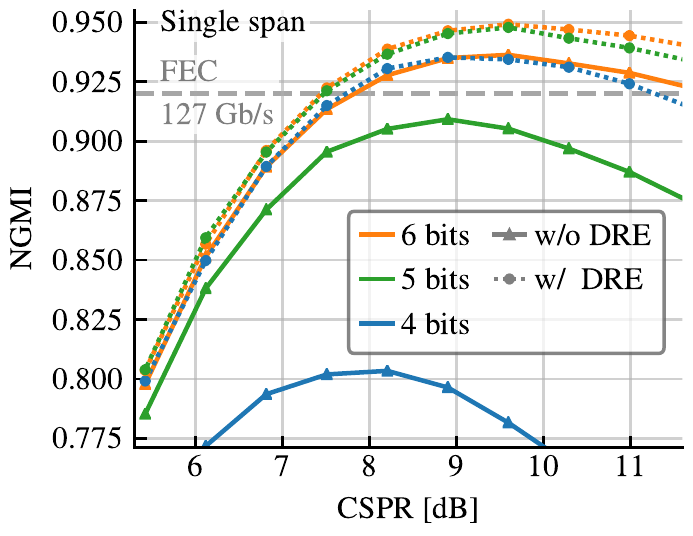}\vspace*{-2mm}
        \caption{}\label{fig:ss_cspr}
    \end{subfigure}%
    \begin{subfigure}[t]{.4\textwidth}
        \centering
        \includegraphics[width=\textwidth]{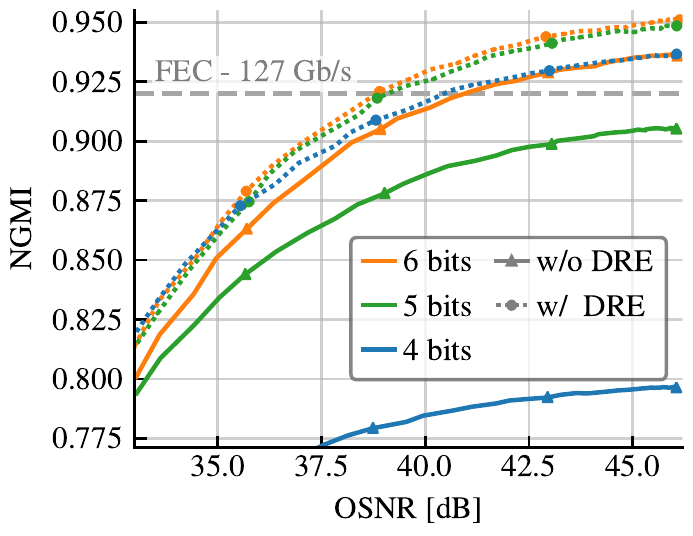}\vspace*{-2mm}
        \caption{}\label{fig:ss_osnr}
    \end{subfigure}%
  \begin{subtable}[b]{0.2\textwidth}
      \centering
      \footnotesize
            \setlength\tabcolsep{1.5pt}
            \begin{tabular}{p{1.0cm}cc}\toprule
                 & w/o DRE & w/ DRE \\\midrule
                \multicolumn{1}{l}{\emph{6 bits}}& & \\
                \multicolumn{1}{r}{NGMI} & 0.936 & 0.949 \\
                \multicolumn{1}{r}{CSPR}&  9.6 & 9.6 \\
                \multicolumn{1}{l}{\emph{5 bits}}& & \\
                \multicolumn{1}{r}{NGMI} & 0.909 & 0.948 \\
                \multicolumn{1}{r}{CSPR}&  8.9 & 9.6 \\
                \multicolumn{1}{l}{\emph{4 bits}}& & \\
                \multicolumn{1}{r}{NGMI} & 0.803 & 0.935 \\
                \multicolumn{1}{r}{CSPR}&  8.2 & 8.9
                \\\bottomrule
            \end{tabular}
      \caption{}
      \label{table:ss}
  \end{subtable}
      \vspace{-5mm}
    \caption{Experimental results for transmission over \SI{50}{\km}, showing in \textbf{\subref{fig:ss_cspr}} the \gls{NGMI} vs \gls{CSPR} and in \textbf{\subref{fig:ss_osnr}} \gls{NGMI} vs \gls{OSNR} at the optimal \gls{CSPR}. \textbf{\subref{table:ss}} summarizes the maximum \gls{NGMI} and optimum \gls{CSPR}.}
\label{fig:ss}\vspace{-6mm}
\end{figure}

\cref{fig:b2b_cspr} shows the results from a \gls{CSPR} sweep performed in \gls{OBTB} configuration for a varying number of bits, with and without \gls{DRE} enabled. A \gls{FEC} limit of 0.92 corresponding to a data rate of \SI{127}{\giga\bit\per\second} is also shown \cite{Alvarado2018}. As can be seen from \cref{fig:b2b_cspr,table:bb}, when using all available 6 bits of the \gls{DAC} system, enabling \gls{DRE} increases the \gls{NGMI} by 0.008. This corresponds to a \SI{1}{\dB} reduction in required \gls{OSNR} at \gls{FEC} limit, as seen from \cref{fig:b2b_osnr}, where the \gls{OSNR} was swept at the optimal \gls{CSPR}. Limiting the available bits of the \gls{DAC} to 5 and 4 bits reduces the performance, but enabling \gls{DRE} increases the performance with 0.027 and 0.11 in \gls{NGMI}, respectively. From \cref{table:bb} it is also observed that the optimal \gls{CSPR} increases when using \gls{DRE}, due to the lower impact of quantization noise.

In \cref{fig:ss}, transmission results over \SI{50}{\km} of fiber are shown. Compared to \gls{OBTB}, the performance is reduced, which can be attributed to the additional dispersion of the \SI{50}{\km} fiber on the \gls{KK} reconstruction. Similar performance gains are obtained by using \gls{DRE}, 0.013 in \gls{NGMI} when using the full 6 bits of the \gls{DAC}, corresponding to a \SI{2}{\dB} reduction in required \gls{OSNR} at \gls{FEC}, see \cref{fig:ss_osnr}. Furthermore, transmitting with only 4 or 5 bits required \gls{DRE} to reach the \gls{FEC} limit. From \cref{table:ss} it is seen that in the transmission scenario, the optimum \gls{CSPR} increases as well while enabling \gls{DRE}.

\section{Conclusions}
We demonstrated the use of \gls{DRE} to increase the performance of a \gls{KK} receiver with the \gls{CW} tone added digitally at the transmitter. Using the full resolution of the \glspl{DAC}, an \gls{NGMI} gain of \num{0.013} was observed for transmission over \SI{50}{\km} of \gls{SSMF}, corresponding to a \SI{2}{\dB} reduction in required \gls{OSNR}. 
Furthermore, we study the performance of \gls{KK} receiver combined with \gls{DRE} at the resolution of 4 and 5 bits. The results indicate the ability to use a \gls{KK} scheme using a digital tone with a \gls{DAC} with a lower number of bits.


\noindent
\emph{\footnotesize{Partial funding is from the KPN-TU/e Smart Two program and from the Dutch NWO Gravitation Program on Research Center for Integrated Nanophotonics (Grant Number 024.002.033).}}
\vspace{-2mm}
\bibliographystyle{style/osajnl}
\bibliography{ref.bib}

\end{document}